\documentclass[11pt,oneside,a4paper]{article}
\usepackage{graphicx}
\usepackage {amsmath}
\usepackage{authblk}
\usepackage{hyperref}
\usepackage{amsthm}
\usepackage{amssymb}
\usepackage{float}
\usepackage[]{blindtext}

\title{Interacting Field Cosmological Model in Lyra Geometry} 

\author[1]{R. V. MAPARI}
\author[2]{D. D. PAWAR}
\author[3]{V. R. PATIL}
\author[4]{J. L. PAWDE}

\affil[1]{Department of Mathematics, Government Science College, Gadchiroli, 442 605 (India)}

\affil[2]{School of Mathematical Sciences, Swami Ramanand Teerth Marathwada University, Nanded, 431 606 (India)}

\affil[3,4]{ Department of Mathematics, Arts, Science \& Commerce College, Chikhaldara, Dist. Amravati, 444 807 (India)}

\date{}

\begin{document}
\maketitle
{{\bf Email:} $^1$r.v.mapari@gmail.com}
\begin{abstract}
The paper explores a plane symmetric cosmological model within the framework of the Lyra manifold, incorporating interactions among various fields. These fields include a charged perfect fluid, a mass-less scalar field, and an electromagnetic field. The study focuses on deriving relativistic field equations and exact solutions for this complex system. The relationships between the scalar field ($\beta$) and the average scale factor ($a(t)$), as well as between the metric potentials, are assumed to solve the nonlinear field equations. Two specific expansion models are examined: 1) Exponential expansion and 2) Power law expansion. The research delves into the dynamic parameters' behavior, observing changes in pressure, density, and cosmological parameters across different models. The findings indicate that while the Universe is expanding, the rate of expansion can vary among the different models. The investigation also involves kinematic parameters, such as jerk and snap parameters derived from the scale factor variation, and compares them to the $\Lambda$CDM model.
\end{abstract}
\section{\NoCaseChange{} Introduction}
\paragraph{}The Supernovae cosmology project and the High-Z Supernova search team made significant observations regarding the late-time acceleration of the Universe. This discovery has captured widespread curiosity and is driving remarkable scientific advancements in cosmology. The increasing attention towards modern cosmology stems from its novel approach to studying the Universe, particularly the concept of accelerated expansion. Numerous scientists have contributed to the observation and exploration of this phenomenon through various studies \cite{d1, d2, d3, d4, d5, d6, d7}. Researchers have found the concept of the Universe's accelerated expansion to be more mysterious than merely beautiful and intriguing. This phenomenon presents a significant challenge in terms of understanding its exact cause. Through dedicated efforts, advancements have been achieved in cosmology, and numerous scientists have shared their insights into this enigma. They propose that an invisible force known as dark energy is responsible for this phenomenon. To comprehend the nature of the accelerated expansion and dark energy, several theoretical models have been put forth, including the quintessence scalar field \cite{d8,d9}, phantom field \cite{d10,d11,d11a}, K-essence \cite{d12,d13,d13a}, tachyon field \cite{d14,d15}, quintom \cite{d16,d17}, and Chaplygin gas \cite{d18,d19}. These models aim to unravel the mysteries behind this remarkable phenomenon. Einstein's theory was crucial for understanding the Universe's origin and evolution, but it couldn't fully explain its later accelerated expansion. To address this, scholars have modified gravity theories to better describe this phase. Multiple alternative gravity theories have been developed and tested to account for the Universe's late-time accelerated expansion.
\paragraph{}In the context of Riemannian geometry modifications, Weyl's work \cite{d20} lacked satisfactory physical significance. Lyra introduced a significant improvement by incorporating a new gauge function into a structure-less manifold \cite{d21}. This approach was later examined by Sen \cite{d22} and Sen along with Dunn \cite{d23}, building upon Lyra's geometry. Sen further formulated field equations for the Lyra manifold using the normal gauge, as discussed in the following section. Halford pointed out that the vector field $\phi_i$ in Lyra's geometry serves a role analogous to the cosmological constant $\Lambda$ in Einstein's theory \cite{d24}. In recent research, there has been a connection proposed between dark energy and the cosmological constant $\Lambda$ \cite{d25}, leading to a fresh perspective on their relationship \cite{d26}. Numerous studies have diligently explored the behavior of the Universe within the context of Lyra geometry, as indicated by references \cite{d27, d28, d29, d30, d31, d32, d33, d34}. These references collectively suggest that within Lyra geometry, the displacement field denoted as $\beta$ functions akin to a cosmological constant, holding a connection with the average scale factor. An important observation is that $\beta$ also exhibits a potential link to dark energy.
\paragraph{}The study focuses on the investigation of a plane symmetric cosmological model and examines the inhomogeneity of the Universe. Previous research on this topic by various researchers has yielded interesting results. Notably, Sahoo and Mishra recently explored the plane symmetric Universe within the context of bimetric relativity, finding it inadequate in explaining the early stages of the Universe. They also analyzed the model within the framework of scale invariant theory, revealing an initial singularity and a consistent shape of the Universe throughout its evolution. Another study by Katore and Shaikh identified a nonsingular and expanding Universe within the plane symmetric cosmological model. Additionally, some authors have investigated the possibility of a radiating Universe within the constraints of the plane symmetric metric. These findings collectively contribute to our understanding of the nature of the Universe in the context of plane symmetry. (for review one can refer: \cite{d35, d36, d37, d38, d39, d40, d41, d42, d43, d44, d45, d46}). Agrawal and Pawar conducted a study focusing on plane symmetric Universes within the context of modified gravity. They analyzed the cosmological behavior of these Universes, and their findings were documented in reference \cite{d47}. In a similar vein, Bayskar and colleagues explored plane symmetric Universes with interacting fields, but within the framework of general relativity, as detailed in reference \cite{d48}. Building upon this, Pawar and Mapari investigated plane symmetric Universes incorporating interacting fields, but within the realm of modified gravity. Their work revealed a transitional phase for the Universe, as discussed in reference \cite{d49}. Other scholars have also delved into the realm of plane symmetric Universes, particularly when influenced by massless scalar fields, as referenced in \cite{d50}. Additionally, Mohanty et al. studied charged stiff fluid coupled with massless scalar fields, as outlined in reference \cite{d51}. Also, Mohanty examines the challenges related to interacting fields in a relativistic cylindrically symmetric universe \cite{d52}. Panigrahi and Sahu \cite{d53} investigate micro and macro cosmological models incorporating a massless scalar field. The $\Lambda CDM$ model is currently favored due to its compatibility with observational data, though it has issues of fine tuning and cosmological coincidence \cite{d54, d55}. The authors introduce jerk and snap parameters, particularly their importance in approximating the $\Lambda CDM$ model (where jerk parameter, $j=1$). A comparison is made between the presented model and the $\Lambda CDM$ model in terms of the evolution of the jerk parameter, discussed in the observation and discussion section.
\paragraph{}The above literature has inspired us to pursue further research in this domain. The paper is structured into distinct sections. Section 2 focuses on metric and field equations, while Section 3 outlines the methodology for solving these field equations. In Section 4, two models are presented: Section 4.1 details the exponential expansion model, and Section 4.2 covers the power law expansion model. Section 5 contains observations and discussions. The conclusions drawn from the study are summarized in Section 6.

\section{The metric and field equations}
We have considered plane symmetric metric of the form,
\begin{equation} \label{e1} 
	ds^2=dt^2-A^2\left(dx^2+dy^2\right)-B^2dz^2 
\end{equation} 
Where, $A=A(t)$ and $B=B(t)$ are metric potentials.

Modification of general relativity (GR) is the Lyra manifold. The field equations for Lyra's manifold are given by (Sen, 1957) are,
\begin{equation} \label{e2} 
	R_{ij}-\frac{1}{2}Rg_{ij}+\frac{3}{2}{\phi}_i{\phi}_j-\frac{3}{4}g_{ij}{\phi }_k{\phi }^k=-{\mathop{T}\limits^{\leftrightarrow}}_{ij} 
\end{equation} 
Where, ${\phi }_i$ are displacement field and we have chosen $8\pi G=c=1$.\\ 
Here
\begin{equation} \label{e3} 
	{\phi }_i=(0,0,0,\beta (t)) 
\end{equation}
Also, displacement field satisfies,
\begin{equation} \label{e4} 
	{\phi }_i{\phi }^j=\left\{ \begin{array}{c}
		0,\ \ for\ i,j=1,2,3 \\ 
		{\ \beta }^2(t),\ \ for\ i,j=4 \end{array}
	\right. 
\end{equation}
We consider the energy-momentum tensor as a representation of an interacting field and undertake an analysis of the characteristics of the cosmological model within the framework of coexisting linearly coupled perfect fluid distribution, a mass-less scalar field, and a source representing a free electromagnetic field. Therefore in Eqn.(\ref{e2}),  $\mathop{T_{ij}}\limits^{\longleftrightarrow}$  is given by,
\begin{equation} \label{e5} 
	\mathop{T_{ij}}\limits^{\longleftrightarrow}=S_{ij}+T_{ij}+E_{ij} 
\end{equation} 
In the Eqn.(\ref{e5}), ${S}_{ij}$ represents the energy source for perfect fluid distribution, $T_{ij}$ is the energy source for mass-less scalar field and $E_{ij}$ represents the electromagnetic energy momentum tensor and it is given by,
\begin{equation} \label{e6} 
	S_{ij}=\left(p+\rho \right)u_iu_j-g_{ij}p 
\end{equation} 
Together with 
\begin{equation} \label{e7} 
	g^{ij}u_iu_j=1 
\end{equation} 
Where $p$ is internal pressure, $\rho$ is rest mass density and $u^i$ is four-velocity vectors of the distribution.
\begin{equation} \label{e8} 
	T_{ij}=U_{,i}U_{,j}-\frac{1}{2}g_{ij}U_{,s}U^{,s} 
\end{equation} 
Where $U$ is Mass-less scalar field. 
\begin{equation} \label{e10} 
	E_{ij}=\frac{1}{4\pi }\left[F_{i\alpha }F^{\alpha }_j-\frac{1}{4}g_{ij}F_{\alpha \beta }F^{\alpha \beta }\right] 
\end{equation} 
Here $F_{ij}$ is the electromagnetic field tensor which is obtained from the four potential ${\phi }_i$,
\begin{equation} \label{e11} 
	F_{ij}={\phi }_{i,j}-{\phi }_{j,i} 
\end{equation} 
\begin{equation} \label{e12} 
	F^{ij}_{;j}=-4\pi {\rho }_cu^i 
\end{equation} 
In a co-moving transformation system the magnetic field is considered along the $z$-axis only, therefore non-vanishing components of electromagnetic fields $F_{ij}$ are only $F_{12}$ and $F_{21}$. Also, we have an electromagnetic field tensor that is anti-symmetric.

The first set of Maxwell's equation,
\begin{equation} \label{e13} 
	F_{ij,k}+F_{jk,i}+F_{ki,j}=0 
\end{equation} 
It gives, \begin{equation} \label{e14} 
	F_{12}=constant=M 
\end{equation}   
Now from Eqn. (\ref{e6}), (\ref{e8}), (\ref{e10}) for the metric Eqn.(\ref{e1}), we have
\begin{equation} \label{e15} 
	E^1_1=E^2_2=-E^3_3=-E^4_4=\frac{M^2}{8\pi A^4} 
\end{equation} 
\begin{equation} \label{e16} 
	T^1_1=T^2_2=T^3_3=-T^4_4=-\frac{{\dot{U}}^2}{2} 
\end{equation} 
\begin{equation} \label{e17} 
	S^1_1=S^2_2=S^3_3=-p\ ;\ \ S^4_4=\rho \  
\end{equation}
By using Eqn.(\ref{e15}) to (\ref{e17}), field Eqn.(\ref{e2}) of Lyra Manifold can be reduced for the Eqn. (\ref{e1}) as follows,
\begin{equation} \label{e18} 
	\frac{\ddot{A}}{A}+\frac{\ddot{B}}{B}+\frac{\dot{A}\dot{B}}{AB}-\frac{3}{4}A^2{\beta }^2=\frac{M^2}{8\pi A^4}-\frac{{\dot{U}}^2}{2}-p 
\end{equation} 
\begin{equation} \label{e19} 
	\frac{{\dot{A}}^2}{A^2}+2\frac{\ddot{A}}{A}-\frac{3}{4}B^2{\beta }^2=-\frac{M^2}{8\pi A^4}-\frac{{\dot{U}}^2}{2}-p 
\end{equation} 
\begin{equation} \label{e20} 
	\frac{{\dot{A}}^2}{A^2}+2\frac{\dot{A}\dot{B}}{AB}-\frac{3}{4}{\beta }^2=-\frac{M^2}{8\pi A^4}+\frac{{\dot{U}}^2}{2}+\rho  
\end{equation} 
Here dot for differentiation w.r.t. time $t$.
For the metric Eqn.(\ref{e1}), the cosmological parameters are defined as follows.\\
Average scale factor,
\begin{equation} \label{e21} 
	a\left(t\right)={(A^2B)}^{{1}/{3}} 
\end{equation} 
Spatial Volume,
\begin{equation} \label{e22} 
	V=A^2B 
\end{equation} 
Directional Hubble parameters,
\begin{equation} \label{e23} 
	H_x=H_y=\frac{\dot{A}}{A},\ \ H_z=\frac{\dot{B}}{B} 
\end{equation} 
Average Hubble Parameter,
\begin{equation} \label{e24} 
	H=\frac{\dot{a}}{a} 
\end{equation} 
Scalar expansion,
\begin{equation} \label{e25} 
	\theta =3H 
\end{equation} 
Shear scalar,
\begin{equation} \label{e26} 
	{\sigma }^2=\frac{1}{2}{\sigma }^{ij}{\sigma }_{ij} 
\end{equation} 
Average anisotropy parameter,
\begin{equation} \label{e27} 
	\Delta =\frac{1}{3}\sum^3_{i=1}{{\left[\frac{H_i-H}{H}\right]}^2} 
\end{equation} 
Where, $H_i$ are the directional Hubble parameters.\\
Deceleration parameter,
\begin{equation} \label{e28} 
	q=-1+\frac{d}{dt}\left(\frac{1}{H}\right) 
\end{equation}
The jerk parameter ($j$) defined and discussed as (one can refer \cite{d56, d57, d58, d59, d60, d61}).
\begin{equation} \label{j1}
	j(t)=\frac{1}{a}\dddot{a}\left[\frac{1}{a}\dot a\right]^{-3}
\end{equation}
Also, snap parameter ($s$) defined as \cite{d59, d62}.
\begin{equation} \label{s1}
	s(t)=\frac{1}{a}\ddddot{a}\left[\frac{1}{a}\dot a\right]^{-4}
\end{equation}
Here, $\dddot a$ and $\ddddot a$ in Eqn.(\ref{j1}) and (\ref{s1}) are the third and fourth derivative of dimensionless scale factor respectively with respect to time $t$.\\
As $j$ and $s$ play a crucial role in cosmic observations. Some scientists have done intensive research on jerk and snap parameters. According to them, it describes the Universe which is close to the $\Lambda CDM$ model \cite{d63, d64}.

\section{Solution of the field equations}
Now, the system of Eqn.(\ref{e18}) to (\ref{e20}) are highly nonlinear differential equations in which we are interested in $A,\ B,\ U,\ \beta ,\ \ p$ and $\rho $. Therefore for the exact solution of the obtained field equations we have considered feasible constraints. 

We have considered the linear relationship between the metric potentials $A$ and $B$
\begin{equation} \label{e29} 
	A=nB,\ \ \ \ \ \ \ \ \ \ \ \ \ \ \ \ \ \ n\neq 0,\ 1 
\end{equation} 
Where $n$ is an arbitrary positive constant.\\
We have explored the solution of the field equations by considering two volumetric expansion laws discussed in \cite{d33, d66, d67}.
\noindent The exponential law and the power law are given by,
\begin{equation} \label{e31} 
	V=e^{3kt} 
\end{equation} 
\begin{equation} \label{e32} 
	V=t^n 
\end{equation} 
Where $k\ and\ n\ $ are positive constants.                                                                      

\subsection{\textbf{Model for Exponential Expansion}}
\paragraph{}Exponential law for expansion volume factor is, 
\[V=A^2B=e^{3kt}=a^3\] 
It gives,                                            
\begin{equation} \label{e36} 
	a=e^{kt} 
\end{equation} 
Now $\beta$ can be obtained from average scale factor which have been discussed by Johri and Sudarshan \cite{d65}.
\begin{equation} \label{e30} 
	\beta ={\beta }_0a^k 
\end{equation}
Where ${\beta }_0$ and $k$ are positive constant.
\begin{equation} \label{e37} 
	\beta ={{\beta }_0e}^{k^2t} 
\end{equation} 
Now, Eqn.(\ref{e37}), Eqn.(\ref{e18}) and Eqn.(\ref{e19}) gives, 
\begin{equation} \label{e38} 
	B={\left[\frac{M^2}{3\pi n^2(1-n^2){{\beta }_0}^2}\right]}^{\frac{1}{6}}e^{-\frac{k^2}{3}t} 
\end{equation} 
\begin{equation} \label{e39} 
	A={\left[\frac{M^2n^4}{3\pi (1-n^2){{\beta }_0}^2}\right]}^{\frac{1}{6}}e^{-\frac{k^2}{3}t} 
\end{equation} 
From Eqn.(\ref{e38}) and Eqn.(\ref{e39}), a metric Eqn.(\ref{e1}) reduced as,
\begin{align} \label{e40}
	ds^2=dt^2-{\left[\frac{M^2n^4}{3\pi (1-n^2){{\beta }_0}^2}\right]}^{\frac{1}{3}}e^{-\frac{2k^2}{3}t}\left(dx^2+dy^2\right)
	-{\left[\frac{M^2}{3\pi n^2(1-n^2){{\beta }_0}^2}\right]}^{\frac{1}{3}}e^{-\frac{{2k}^2}{3}t}dz^2 
\end{align}	
Also, the parameters defined in Eqn.(\ref{e24}) to (\ref{e28}) are obtained as,
\begin{equation} \label{e42} 
	H=k 
\end{equation} 
\begin{equation} \label{e43} 
	\theta =3k 
\end{equation} 
\begin{equation} \label{e44} 
	{\sigma }^2=0 
\end{equation} 
\begin{equation} \label{e45} 
	\Delta ={\left[\frac{k+3}{3}\right]}^2 
\end{equation} 
\begin{equation} \label{e46} 
	q=-1 
\end{equation} 

\noindent Pressure and density for the exponential expansion model is,

\noindent From,Eqn.(\ref{e18}), 
\begin{align}\label{e47}
	p=&-\frac{k^4}{3}+\frac{3}{4}\left[\frac{M^2\ {{\beta }_0}^4}{3\pi n^2(1-n^2)}\right]^{\frac{1}{3}}e^{\frac{4k^2}{3}t}-\frac{M^2}{8\pi n^4}{\left[\frac{M^2\ }{3\pi n^2(1-n^2){{\beta }_0}^2}\right]}^{-\frac{1}{3}}e^{\frac{2k^2}{3}t}\\ \nonumber
	&-\frac{{c_1}^2M^2}{6\pi n^6(1-n^2){{\beta }_0}^2}e^{-2k^2t}
\end{align}	 
From Eqn. (\ref{e20}), 
\begin{align} \label{e48}
	\rho =&\frac{k^4}{3}+\frac{3}{4}{{\beta }_0}^2e^{2k^2t}+\frac{M^2}{8\pi n^4}{\left[\frac{M^2\ }{3\pi n^2(1-n^2){{\beta }_0}^2}\right]}^{-\frac{2}{3}}e^{\frac{4k^2}{3}t}\\ \nonumber
	&-\frac{{c_1}^2M^2}{6\pi n^6(1-n^2){{\beta }_0}^2}e^{-2k^2t}  
\end{align}
\noindent Now, for the metric Eqn.(\ref{e40}), massless scalar field obtained from Eqn.(\ref{e18}) as,
\begin{equation} \label{e41} 
	U=\frac{c_1k^2}{n^2}{\left[\frac{M^2}{3\pi n^2(1-n^2){{\beta }_0}^2}\right]}^{-\frac{1}{2}}e^{k^2t} 
\end{equation} 
The jerk and snap parameters for the exponential expansion model from Eqn.(\ref{j1}) and Eqn.(\ref{s1}) respectively are,
\begin{equation} \label{j2}
	j(t)=1
\end{equation}
\begin{equation} \label{s2}
	s(t)=1
\end{equation}
It is noteworthy that the present exponential expansion model represents the $\Lambda CDM$ model.  
\begin{figure}[!t]
	\centering
	\includegraphics[width=1\columnwidth]{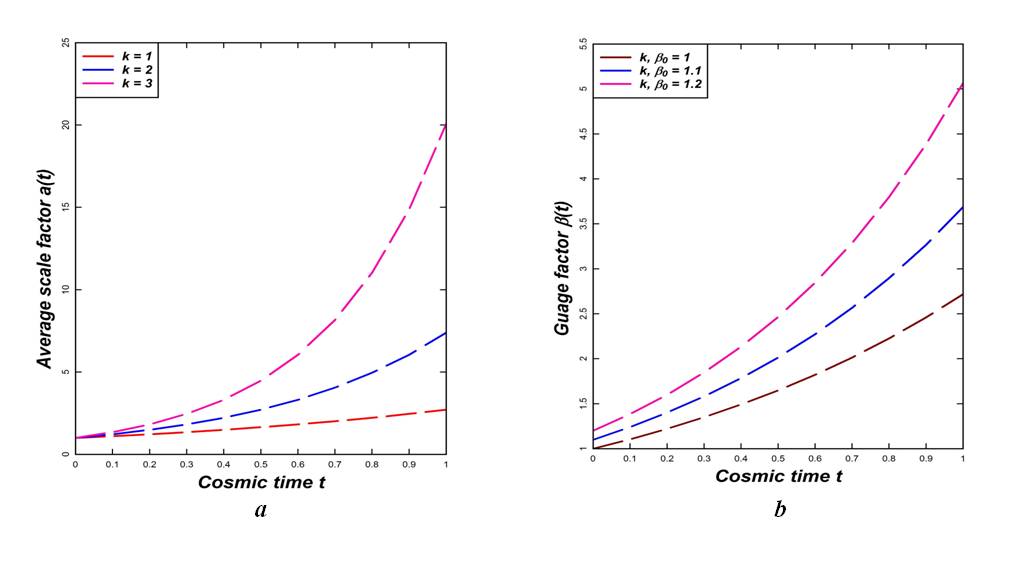}
	\caption{Variation of ($a$) Average scale factor (b) Gauge factor against cosmic time $t$ in Exponential expansion model}
	\label{fige1}
\end{figure}
\begin{figure}[!t]
	\centering
	\includegraphics[width=1\columnwidth]{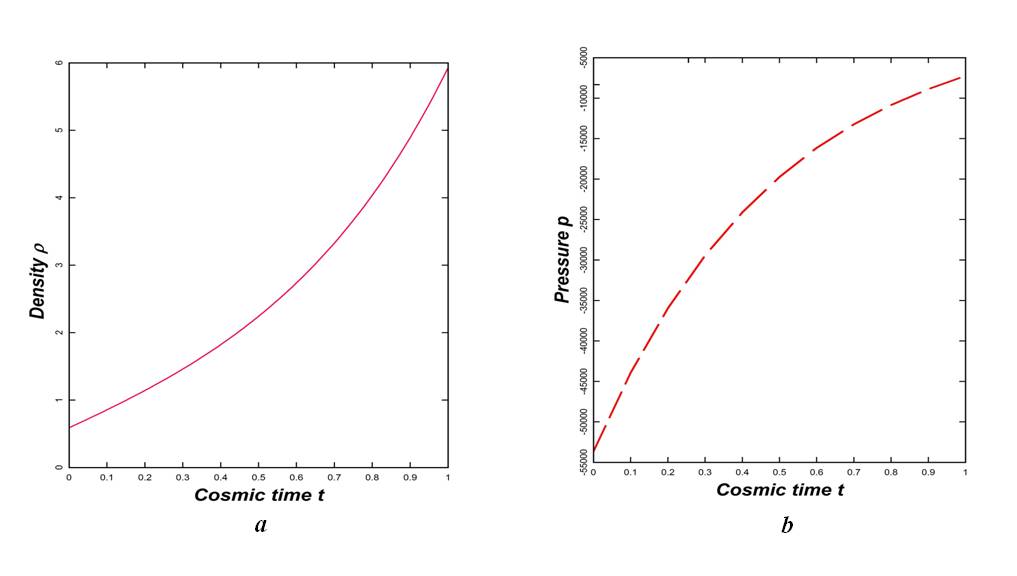}
	\caption{Variation of ($a$) Density (b) Pressure against cosmic time $t$ in Exponential expansion model}
	\label{fige2}
\end{figure}
\subsection{\textbf{Model for Power law Expansion}}
\paragraph{}Power law volumetric expansion is given by, 
\[V=A^2B=t^n=a^3\] 
It gives,
\begin{equation} \label{e49} 
	a=t^{\frac{n}{3}} 
\end{equation} 
\begin{equation} \label{e50} 
	\beta ={{\beta }_0t}^{\frac{nk}{3}} 
\end{equation} 
Now, Eqn.(\ref{e37}), Eqn.(\ref{e18}) and Eqn.(\ref{e19}) gives,
\begin{equation} \label{e51} 
	B={\left[\frac{M^2}{3\pi n^2(1-n^2){{\beta }_0}^2}\right]}^{\frac{1}{6}}t^{-\frac{nk}{9}} 
\end{equation} 
\begin{equation} \label{e52} 
	A={\left[\frac{M^2n^4}{3\pi (1-n^2){{\beta }_0}^2}\right]}^{\frac{1}{6}}t^{-\frac{nk}{9}} 
\end{equation} 
From Eqn.(\ref{e51}) and Eqn.(\ref{e52}), metric Eqn.(\ref{e1}) reduced as,
\begin{align} \label{e53} 
	ds^2=&dt^2-{\left[\frac{M^2n^4}{3\pi (1-n^2){{\beta }_0}^2}\right]}^{\frac{1}{3}}t^{-\frac{nk}{9}}\left(dx^2+dy^2\right)\\ \nonumber
	&-{\left[\frac{M^2}{3\pi n^2(1-n^2){{\beta }_0}^2}\right]}^{\frac{1}{3}}t^{-\frac{nk}{9}}dz^2 
\end{align} 
Also, the parameters defined in Eqn.(\ref{e24}) to Eqn.(\ref{e28}) are obtained as,
\begin{equation} \label{e55} 
	H=\frac{n}{3t} 
\end{equation} 
\begin{equation} \label{e56} 
	\theta =\frac{n}{t} 
\end{equation} 
\begin{equation} \label{e57} 
	{\sigma }^2=0 
\end{equation} 
\begin{equation} \label{e58} 
	\Delta ={\left[\frac{k+3}{3}\right]}^2 
\end{equation} 
\begin{equation} \label{e59} 
	q=\frac{3}{n}-1 
\end{equation} 
\noindent Pressure and density for the power law expansion model is,

\noindent From,Eqn.(\ref{e18}),
\begin{align} \label{e60} 
	p=&-\frac{nk}{9}\left(\frac{nk}{3}+2\right)t^{-2}+\frac{3}{4}\left[\frac{M^2{\beta_0}^4}{3\pi  n^2(1-n^2)}\right]^{\frac{1}{3}}t^{\frac{4nk}{9}}\\ \nonumber
	&-\frac{M^2}{8\pi n^4}{\left[\frac{M^2\ }{3\pi n^2(1-n^2){{\beta }_0}^2}\right]}^{-\frac{1}{3}}t^{\frac{2nk}{9}}-\frac{{c_1}^2M^2}{6\pi n^6(1-n^2){{\beta }_0}^2}t^{-\frac{2nk}{3}} 
\end{align} 
From Eqn. (\ref{e20}), 
\begin{align} \label{e61} 
	\rho =&\frac{n^2k^2}{27}t^{-2}-\frac{3}{4}{{\beta }_0}^2t^{\frac{2nk}{3}}+\frac{M^2}{8\pi n^4}{\left[\frac{M^2\ }{3\pi n^2(1-n^2){{\beta }_0}^2}\right]}^{-\frac{2}{3}}t^{\frac{4nk}{9}}\\ \nonumber
	&-\frac{{c_1}^2M^2}{6\pi n^6(1-n^2){{\beta }_0}^2}t^{-\frac{2nk}{3}}
\end{align} 
\noindent Now, for the metric Eqn.(\ref{e53}), From Eqn.(\ref{e18}) massless scalar field is,
\begin{equation} \label{e54} 
	U=\frac{c_1k}{3n}{\left[\frac{M^2}{3\pi n^2(1-n^2){{\beta }_0}^2}\right]}^{-\frac{1}{2}}t^{\frac{nk}{3}-1} 
\end{equation}
The jerk and snap parameters for the power law expansion model obtained from Eqn.(\ref{j1}) and Eqn.(\ref{s1}) respectively are,
\begin{equation} \label{j3}
	j(t)=\frac{9}{n^2}\left(\frac{n}{3}-1\right)\left(\frac{n}{3}-2\right)
\end{equation}
\begin{equation} \label{s3}
	s(t)=\frac{9}{n^2}\left(\frac{n}{3}-1\right)\left(\frac{n}{3}-2\right)\left(\frac{n}{3}-3\right)
\end{equation}
The present power law expansion model found close to $\Lambda CDM$ model. 
\begin{figure}[!t]
	\centering
	\includegraphics[width=1\columnwidth]{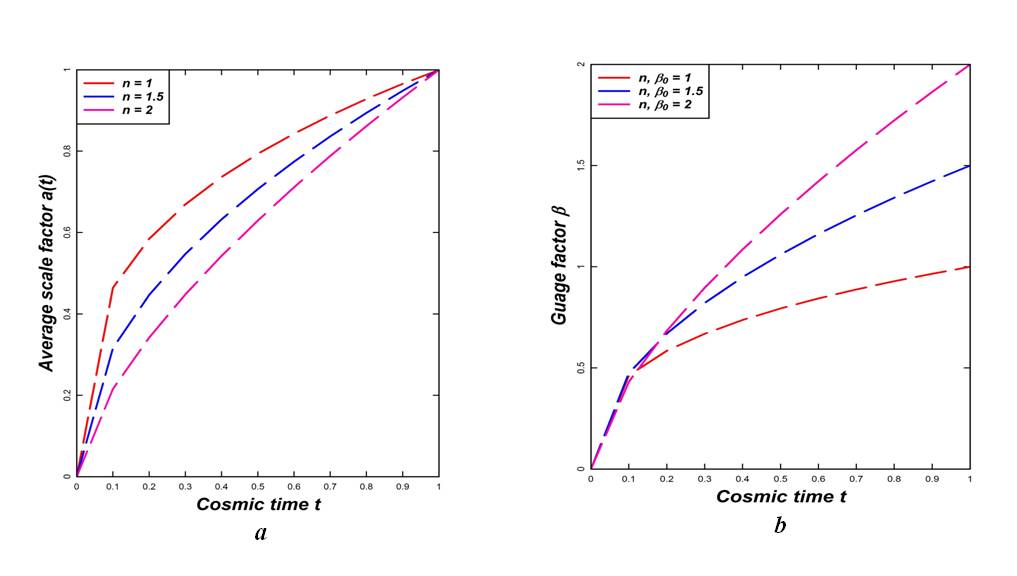}
	\caption{Variation of ($a$) Average Scale Factor (b) Gauge factor against cosmic time $t$ in Power law expansion model}
	\label{figp1}
\end{figure}
\begin{figure}[!t]
	\centering
	\includegraphics[width=1\columnwidth]{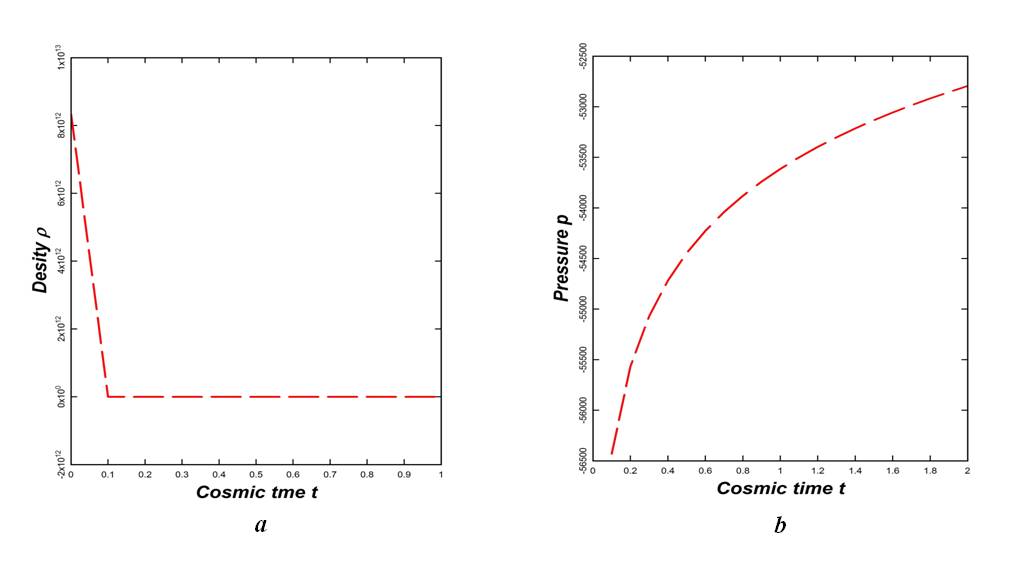}
	\caption{Variation of ($a$) Density (b) Pressure against cosmic time $t$ in Power law expansion model}
	\label{figp2}
\end{figure}
\begin{figure}[!t]
	\centering
	\includegraphics[width=1\columnwidth]{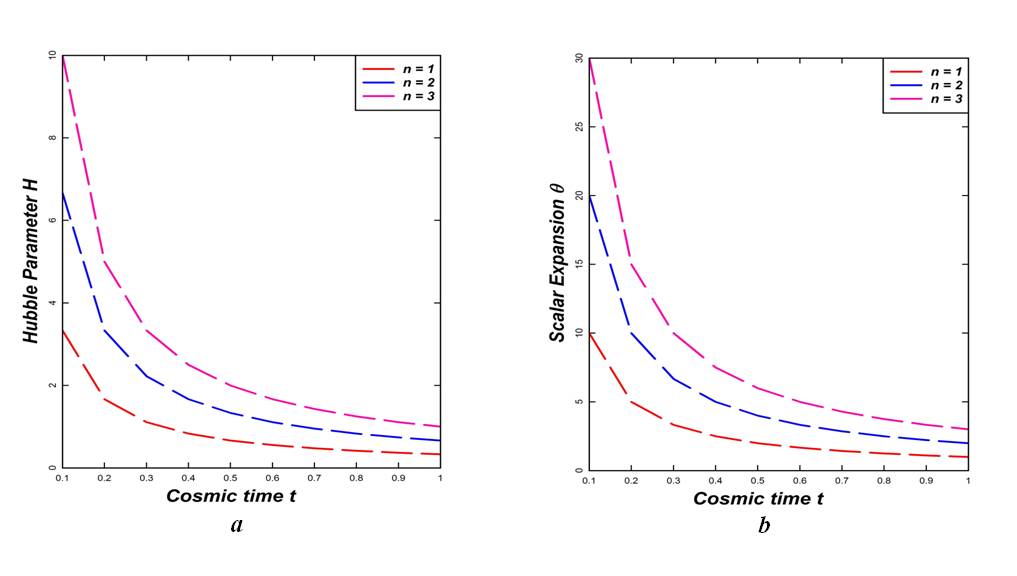}
	\caption{Variation of ($a$) Hubble parameter (b) Scalar expansion against cosmic time $t$ in Power law expansion model}
	\label{figp3}
\end{figure}
\begin{figure}[!t]
	\centering
	\includegraphics[width=1\columnwidth]{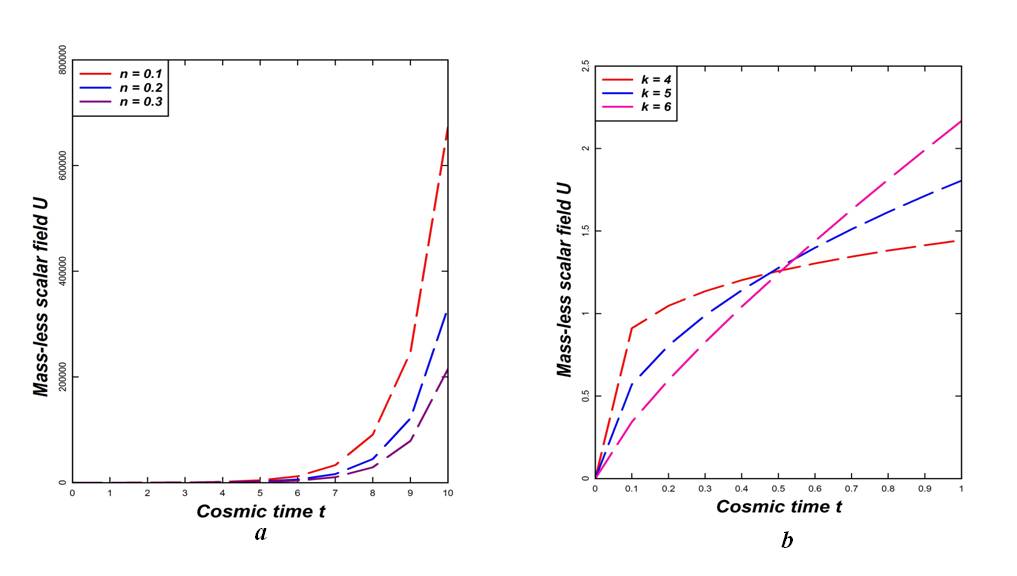}
	\caption{Variation of Massless scalar field in ($a$) Exponential expansion model (b) Power law expansion model  against cosmic time $t$}
	\label{figep1}
\end{figure}
\section{Observations and discussion}
Observations and discussion as follows, 
\begin{itemize}
	\item In cosmology, the cosmic scale factor denoted as $a(t)$ holds significant importance due to its central role in the Friedmann equations. Through our analysis, we've established that $a(t)$ consistently grows with cosmic time ($t$) in both our model (Figure \ref{fige1}(a) and Figure \ref*{figp1}(a)). Our findings align with those presented in a previous study (cite: [d68]). By examining the derivatives of $a(t)$, we've been able to ascertain the jerk and snap parameters. A positive value for the second-order derivative indicates that the rate of expansion of the Universe, represented by $\dot a(t)$, is accelerating.
	\item The displacement field's gauge factor, denoted as $\phi_i$, serves a role similar to the varying cosmological constant $\Lambda (t)$, which is associated with dark energy sources according to references \cite{d9} and \cite{d26}. In this study, a time-dependent gauge factor denoted as $\beta (t)$ was identified, aligning with the findings of various authors and in agreement with references \cite{d69, d70, d71, d72, d73}. Notably, the gauge factor in the proposed model exhibits a consistent increase over cosmic time $t$, with a tendency to approach infinity as $t$ approaches infinity. This behavior holds true for both models presented in Figure.\ref{fige1}(b) and Figure.\ref{figp1}(b).	
	\item In the context of cosmic models, two expansion scenarios are compared: the exponential expansion model and the power law expansion model. In the exponential model (Figure \ref{fige2}($a$)), the density starts with a finite value at the beginning and grows as cosmic time $t$ increases. On the other hand, in the power law model (Figure \ref{figp2}($a$)), the density is infinite at the initial moment and subsequently diminishes as time progresses. These distinct behaviors highlight the differences between the two expansion models.
	\item The identification of negative internal pressure implies the existence of dark energy in the Universe. This conclusion is drawn from our examination of both exponential and power law expansion models, where we note a transition in pressure from a significantly negative magnitude to a less negative one (Figure.\ref{fige2}(b) and Figure.\ref{figp2}(b)). These findings are consistent with the outcomes reported in prior research \cite{d66, d74}.
	\item In the context discussed, the direction of HP's sign ($H<0$ or $H>0$) signifies whether the Universe is contracting or expanding. The scalar expansion value $\theta$ quantifies the rate of expansion. In the current model under consideration, both discussed scenarios exhibit an expansion ($H>0$), but they display distinct characteristics.
	\item In the exponential expansion model, the parameters $H$ and $\theta$ are determined as $k > 0$ and $\theta = 3k$, respectively. This implies a constant rate of expansion for the Universe. Importantly, this model doesn't exhibit an initial singularity.
	\item In the context of the power law expansion model, it was observed that both the Hubble Parameter ($H$) and the expansion rate representative ($\theta$) are dependent on cosmic time ($t$). A significant supporting point for the Big Bang theory is that throughout various time intervals in cosmic history ($t$), both $H$ and $\theta$ are consistently greater than zero ($H > 0$ and $\theta > 0$). Notably, at the very beginning of cosmic time ($t = 0$), both $H$ and $\theta$ take on infinite values, as indicated by equations (\ref{e55}) and (\ref{e56}), and further demonstrated in Figure \ref{figp3}. This implies that the power law expansion model exhibits a singularity at the initial moment of time ($t = 0$).
	\item The interaction between the cosmological constant ($\Lambda$) and a massless scalar field ($U$) was explored by Sahu in reference \cite{d75}. Sahu derived time-dependent expressions for the scalar field $U$ in two different models. Although both models yielded similar observations regarding the time dependence of $U$, there was a discrepancy concerning the convergence and divergence behavior of $U$. In the current study, it was found that the scalar field $U$ converges as time ($t$) approaches $0$, but diverges as time approaches infinity (as depicted in Figure [\ref{figep1}($a$) and \ref{figep1}(b)]). Importantly, it was highlighted that the scalar field $U$ is not defined when $n=1$.
	\item In the context of cosmological expansion models, the deceleration parameter ($q$) is used to describe the acceleration or deceleration of the Universe's expansion. A value of $q=-1$ signifies exponential expansion, while for the power law expansion model, $q<0$ when the parameter $n>3$, indicating accelerated expansion. The findings align well with those presented in references \cite{d67, d76}.
\end{itemize}
\section{Conclusion}
We have studied a new class of interacting field cosmological model in the framework of Lyra manifold. The study focused on an interacting field model and derived various physical parameters to gain insights into the Universe's behavior. The significance of the jerk and snap parameters in understanding the Universe's evolution was highlighted, particularly as a means to explore models akin to the $\Lambda$CDM model. Notably, the findings indicated that the exponential expansion model exhibited a constant jerk value ($j=1$), aligning with the characteristics of the flat $\Lambda$CDM model.The power law expansion model, described by Eqn.(\ref{j3}), leads to a flat $\Lambda$CDM model when the parameter $n=2$. When $n\rightarrow 2^-$, the parameter $j\rightarrow1+\epsilon$, which exhibits characteristics resembling the departures from the $\Lambda$CDM model as observed in references \cite{d63, d77, d78}. Conversely, as $n$ approaches slightly greater than 2, the parameter $j$ approaches a value of $1-\epsilon$, causing the model to approach the traditional $\Lambda$CDM model.In this power law expansion model, when $n$ takes the value of 2, the parameter $s=-\frac{7}{2}$. This value of $s=-\frac{7}{2}$ aligns with the predictions of the $\Lambda$CDM model, as it mirrors the transition observed from $s=-\frac{7}{2}$ to $s=-2$ in the flat $\Lambda$CDM model. This is noteworthy since Poplawski's work in $f(R)$ gravity, as cited in reference \cite{d79}, yielded a value of $s=-2.68$.In the context of the Big Bang theory, compelling evidence has been gathered regarding the behavior of the cosmos. The current model being discussed, which involves exponential and power law expansion, indicates that certain cosmological parameters, specifically $H>0$ and $q<0$, align with the observed expansion and acceleration of the Universe. This alignment with observations has been supported by numerous researchers \cite{d1, d2, d3, d4, d80}.It's important to note that the rate of expansion differs between the two models. In the initial stages, as indicated by Eqn.(\ref{e55}) and Eqn.(\ref{e56}), the values $H\rightarrow \infty$ and $\theta\rightarrow \infty$, thereby confirming the occurrence of the Big Bang within the power law expansion model. Additionally, the presence of dark energy is suggested by negative pressure and a positive value for the parameter $\beta$ in the scalar field $\phi_i$.

\end{document}